\documentstyle{article}

\def\be{\begin{equation}}

\def\bea{{\begin{eqnarray}}}
\def\eea{{\end{eqnarray}}}

\def\a{{\alpha}}
\def\b{{\beta}}
\def\c{{\gamma}}
\def\d{{\delta}}
\def\e{{\epsilon}}
\def\D{{\cal D}}
\def\s{\sigma}
\def\half{{1 \over 2}}
\def\ra{{\rangle}}
\def\la{{\langle}}

\def\ih{{i \over \hbar}}

\def\E{{\cal E}}
\def\au{\underline \alpha}
\def\p{{\bf p }}

\def\x{{\bf x}}

\def\bx{{\bar x}}

\def\ria{{\rightarrow}}
\def\Tr{{\rm Tr}}

\def\sp{{\sigma_{\!{+}}}}
\def\sm{{\sigma_{\!{-}}}}

\begin{document}

\title{Decoherent Histories for Spacetime Domains}

%\titlerunning{Decoherent Histories for Spacetime Domains}

\author{J.J.Halliwell\\
Blackett Laboratory \\ Imperial College \\ London, SW7 2BZ \\ UK}

\maketitle              % typesets the title of the contribution

\begin{abstract}
The decoherent histories approach is a natural medium in which to
address problems in quantum theory which involve time in a non-trivial
way. This article reviews the various attempts and difficulties involved
in using the decoherent histories approach to calculate the probability
for crossing the surface $x=0$ during a finite interval of time. The commonly
encountered difficulties in assigning crossing times arise here as
difficulties in satisfying the consistency (no-interference) condition.
This can be overcome by introducing an environment to produce decoherence,
and probabilities exhibiting the expected classical limit are obtained.
The probabilities are, however, dependent to some degree on the
decohering environment. The results
are compared with a recently proposed irreversible detector model.
A third method is introduced, involving continuous quantum measurement
theory. Some closely related work on the interpretation of the
wave function in quantum cosmology is described.
\end{abstract}

%\documentstyle{article}

%\input defs.tex

%\begin{document}

\section{Introduction}

Although opinions differ as to the value and achievements of
attempts to quantize the gravity, it is undeniable that this
endeavour has inspired a considerable amount of work in a variety
of related fields. In particular, the quantization of gravity puts
considerable pressure on both the mathematical and conceptual
foundations of quantum theory, so it is perhaps not surprising
that many researchers in quantum gravity have been drawn into
working on the foundations of quantum mechanics.

One of the key issues that arises in
the quantization of gravity is the ``Problem of Time''. In the
quantization of cosmological models, the wave function of the
universe satisfies not a Schr\"odinger equation, but the
Wheeler--DeWitt equation,
\begin{equation}
{\cal H} \Psi [h_{ij}, \phi ] = 0
\label{a1}
\end{equation}
The wave function $\Psi$ depends on the three-metric $h_{ij}$
and the matter field configurations $\phi$ on a closed spacelike
three-surface \cite{Har1,Har3,Hal4}. There is no time label. Its
absence is deeply entwined with the four-dimensional
diffeomorphism invariance of general relativity. It is often
conjectured that  ``time'' is somehow already present amongst the
dynamical variables $h_{ij}, \phi $, although to date it has
proved impossible to extract a unique, globally defined time
variable.

Although a comprehensive scheme for interpreting the wave function
is yet to be put forward, a prevalent view is that the
interpretation will involve treating all the dynamical variables
$h_{ij}, \phi$ on an equal footing, rather than trying to single
out one particular combination of them to act as time. For this
reason, it is of interest to see if one can carry out a similar
exercise in non-relativistic quantum mechanics. That is, to see
what the predictions quantum mechanics makes about {\it spacetime}
regions, rather than regions of space at fixed moments of time.

Such predictions are not the ones that quantum mechanics usually makes.
In standard non-relativistic quantum mechanics, the probability of finding a
particle between points $x $ and $ x + dx $ at a fixed time $ t $ is given
by
\begin{equation}
p(x,t) dx = | \Psi (x,t ) |^2 \ dx
\label{a2}
\end{equation}
where $\Psi( x,t)$ is the wave function of the particle.
More
generally, the variety of questions one might ask about a particle
at a fixed moment of time may be represented by a projection
operator $P_{\a}$, which is exhaustive
\begin{equation}
\sum_{\a} P_{\a} = 1
\label{a3}
\end{equation}
and mutually exclusive
\begin{equation}
P_{\a} P_{\b} = \delta_{\a \b} P_{\a}
\label{a4}
\end{equation}
The projection operator appropriate to asking questions about
position is $ P = | x \ra \la  x |$. The probability of a
particular alternative is given by
\begin{equation}
p(\a ) = \Tr \left( P_{\a} \rho \right)
\label{a5}
\end{equation}
where $\rho$ is the density operator of the system at the time in question.

The key feature of the above standard formulae is that they do not
treat space and time on an equal footing. Suppose one asks, for example,
the same sort of question with space and time interchanged. That is,
what is the probability of finding the particle at point $x$ in
the time interval $t$ to $t + dt$? The point is that the answer is
{\it not} given by $ | \Psi (x,t) |^2 dt $. The reason for this is
that, unlike the value of $x$ at fixed $t$, the value of $t$ at
fixed $x$ does not refer to an exclusive set of alternatives.
The position of a particle at fixed time is a well-defined quantity
in quantum mechanics, but the time at which a particle is found
at a fixed position is much more difficult to define because
of the possibility of multiple crossings.

This question is clearly a physically relevant one since time is
measured by physical devices which are generally limited in their
precision. It is therefore never possible to say that a physical
event occurs at a precise value of time, only that it occurs in some
range of times. Furthermore, there has been considerable recent
experimental and theoretical interest in the question of tunneling
times \cite{HaS,Lan}. This is the question, given that a particle
has tunneled through a barrier region, how much time did it spend
inside the barrier?

Spacetime questions tend to be rather non-trivial. As
stressed by Hartle, who has carried out a number of investigations
in this area \cite{Har2,Har0,Har},  time plays a ``peculiar and
central role'' in non-relativistic quantum mechanics. It is not
represented by a self-adjoint operator and there appears to be
no obstruction
to assuming that it may be measured with arbitrary precision. It
enters the Schr\"odinger equation as an external parameter. As such,
it is perhaps best thought of as a label referring to a classical,
external measuring device, rather than as a fundamental quantum
observable. Yet time is  measured by physical systems, and all
physical systems are believed to be subject to the laws of quantum
theory.

Given these features, means more elaborate that those usually
employed are required to define quantum-mechanical probabilities
that do not refer to a specific moment of time, and the issue has a
long history \cite{Time,All}. One may find in the literature a variety
of attempts to define questions of time in a quantum--mechanical
way. These include attempts to define time operators
\cite{GRT,Hol,Per}, the use of internal physical clocks
\cite{Har2,Har0} and path integral approaches
\cite{Fer,YaT,Har,Kum}. The literature on tunneling times is a
particularly rich source of ideas on this topic \cite{HaS}.  Many
of these attempts also tie in with the time--energy uncertainty
relations \cite{MaT,KoA}. For a nice review of many of these issues,
see Ref.\cite{Muga}.

This article is concerned with the attempts to solve problems of a
spacetime nature using the decoherent histories approach to
quantum theory \cite{GeH1,GeH2,Gri,Omn}. It is perhaps of interest
to note that, in addition to inspiring work on the question
involving time, considerations of quantum gravity were also partly
responsible for the development of the decoherent histories
approach. For our purposes, the particular attraction of this
approach is that it addresses directly the notion of a ``history''
or ``trajectory'' and in particular shows how to assign
probabilities to them. It is therefore very suited to the question
of spacetime probabilities considered here. This is because the
question of whether a particle did or did not enter a given region
at {\it any} time in a given time interval clearly cannot be
reduced to a question about the state of the particle at a fixed
moment of time, but depends on the entire history of the system
during that time interval.

The decoherent histories approach, for spacetime questions, turns
out to be most clearly formulated in terms of path integrals over
paths in configuration space \cite{YaT,Har,Har3}. The desired
spacetime amplitudes are obtained by summing $\exp\left(\ih S[x(t)]\right)$,
where $S[x(t)]$ is the action, over paths $x(t)$ passing through
the spacetime region in question, and consistent with the initial
state. The probabilities are obtained by squaring the amplitudes
in the usual way.  (The decoherent histories approach is not
inextricably tied to path integrals, however. Operator approaches
to the same questions are also available, but are often more
cumbersome.)

When computed according to the path integral scheme outlined above,
the probability of entering a spacetime region added to the
probability of not entering that region is not equal to $1$, in
general.
This is because of interference. The question of whether a
particle enters a spacetime region, when carefully broken down, is
actually a quite complicated combination of questions about the
positions of the particle at a sequence of times. It is therefore,
in essence, a complicated combination of double slit situations. Not
surprisingly, there is therefore interference and probabilities
cannot be assigned.

From the point of view of the decoherent histories approach to
quantum theory, therefore, the probability of entering a spacetime
region is quite simply {\it not defined} in general for a simple
point particle system, due to the presence of interference. It is
here that the decoherent histories approach, like all the other
approaches to defining time in quantum theory, runs up against its
own particular brand of difficulties.

It is, however, a common feature of the decoherent histories
approach that most of the histories of interest cannot be defined
due to interference -- histories defined by position at more than
one time for example. It is well known that the interference may
be removed by coupling to environment, typically a bath of
harmonic oscillators in a thermal state. We will therefore
consider the above spacetime problem in the presence of an
environment.

The decoherent histories
approach is reviewed in Section 2 and its application to simple
spacetime questions is discussed in Section 3. The inclusion
of the environment to induce decoherence is described in Section 4.

The probabilities produced by the decoherent histories approach
are in some sense somewhat abstract since they do not refer to a
particular measuring device. In Section 5 we therefore introduce a
model measuring device for the purposes of comparison. The
decoherence model of Section 4 consists of quite a crude
environment, which has, however, been very successful in producing
decoherence and emergent classicality. The measurements it
effectively carries out are of a rather robust and crucially,
irreversible, nature. Hence the most important sort of comparison
is with an irreversible detector model. Interestingly, most of the
arrival time models discussed in the literature are not of this
type. It is therefore of interest to develop a model detector, not
dissimilar to the decoherence model, but sufficiently modified to
carry out a more precise measurement.
The comparison between the decoherent histories approach and the
detector model is then carried out in Section 6. This also leads
to the introduction of a third candidate for the crossing time
probability, derived from continuous quantum measurement theory.

In Section 7 we briefly discuss another type of non-trivial time
question, namely, given that a system is in an energy eigenstate,
what is the probability that it will pass through a given region
in configuration space at {\it any} time? The reason this is of
interest is that it is, in essence, the question one needs to
answer in order to interpret solutions to the Wheeler-DeWitt
equation (\ref{a1}).

We summarize and conclude in Section 8.

%\end{document}

%%%%%%%%
%\documentstyle{article}

%\input defs.tex

%\begin{document}

\section{Decoherent Histories Approach to Quantum Theory}

In this Section we give a brief summary of the decoherent histories
approach to quantum theory. It has been described in considerable depth in
many other places \cite{GeH1,GeH2,Gri,Hal1,Hal2,Har3,Ish,Omn,DoH}.

In quantum mechanics, propositions
about the attributes of a system at a fixed moment of time are
represented by sets of projections operators. The projection
operators $P_{\a}$ effect a partition of the possible alternatives
$\a$ a system may exhibit at each moment of time. They are
exhaustive and exclusive, as noted in Eqs.(\ref{a3}), (\ref{a4}).
A projector is said to be {\it fine-grained} if it is of the form
$ | \a \ra \la \a | $, where $\{| \a \ra \}$ are a complete set of
states. Otherwise it is {\it coarse-grained}. A quantum-mechanical
history (strictly, a {\it homogeneous} history \cite{Ish}) is
characterized by a string of time-dependent projections,
$P_{\a_1}^1(t_1), \cdots P_{\a_n}^n(t_n)$, together with an
initial state $\rho$. The time-dependent projections are related
to the time-independent ones by
\begin{equation}
P^k_{\a_k}(t_k) = e^{i H(t_k-t_0)} P^k_{\a_k} e^{-i H(t_k-t_0)}
\end{equation}
where $H$ is the Hamiltonian.
The candidate probability for these homogeneous histories is
\begin{equation}
p(\a_1, \a_2, \cdots \a_n) = {\rm Tr} \left( P_{\a_n}^n(t_n)\cdots
P_{\a_1}^1(t_1)
\rho P_{\a_1}^1 (t_1) \cdots P_{\a_n}^n (t_n) \right)
\label{b1}
\end{equation}
It is straightforward to show that (\ref{b1}) is both non-negative and
normalized to unity when summed over $\a_1, \cdots \a_n$.
However,(\ref{b1}) does not satisfy all the axioms of probability theory, and for
that reason it is referred to as a candidate probability. It does
not satisfy the requirement
of additivity on disjoint regions of sample space. More precisely,
for each set of histories, one may construct coarser-grained
histories by grouping the histories together. This may be achieved,
for example, by summing over the projections at each moment of time,
\begin{equation}
{\bar P}_{{\bar \a}} = \sum_{\a \in {\bar \a} } P_{\a}
\end{equation}
(although this is not the most general type of coarse graining
-- see below).
The additivity requirement is then that the probabilities for each
coarser-grained history should be the sum of the probabilities of
the finer-grained histories of which it is comprised.
Quantum-mechanical interference generally prevents this requirement
from being satisfied. Histories of closed quantum systems
cannot in general be assigned probabilities.

There are, however, certain types of histories for which
interference is negligible, and the candidate probabilities for histories
do satisfy the sum rules.
These histories may be found using the decoherence functional:
\begin{equation}
D({\underline {\a}} , {\underline {\a}'} ) =
\Tr \left( P_{\a_n}^n(t_n)\cdots
P_{\a_1}^1(t_1)
\rho P_{\a_1'}^1 (t_1) \cdots P_{\a_n'}^n (t_n) \right)
\end{equation}
Here $ {\underline {\a}} $ denotes the string $\a_1, \a_2, \cdots
\a_n$. Intuitively, the decoherence functional measures the amount
of interference between pairs of histories.
It may be shown that the additivity requirement is satisfied for all
coarse-grainings if and only if
\begin{equation}
{\rm Re} D({\underline {\a}} , {\underline {\a}'} ) = 0
\label{b2}
\end{equation}
for all distinct pairs of
histories ${\underline {\a}}, {\underline {\a}'}$ \cite{Gri}.
Such sets of histories are said to be {\it consistent}, or
{\it weakly decoherent.}
The consistency condition (\ref{b2}) is typically satisfied only
for coarse--grained histories, and this then often leads
to satisfaction of the stronger condition of {\it decoherence}
\begin{equation}
D(\au, \au' ) = 0
\end{equation}
for $\au \ne \au' $.
The condition of decoherence is associated with the existence of
so-called generalized records. This means that it is possible to
add a projector $R_{\beta}$ at the end of the chain such that
decoherence is preserved and such that the label $\beta$ is perfectly
correlated with the history alternatives $\a_1, \cdots \a_n$.
There is therefore
in principle some physical measurement that could be carried
out at the end of the history from which complete information
about the entire history can be recovered \cite{GeH2,GeH4,Hal5}.

For histories characterized by projections onto ranges of position
at different times, the decoherence functional may be represented
by a path integral:
\begin{equation}
D(\a, \a') = \int_{\a} {\cal D} x \ \int_{\a'} {\cal D} y
\ \exp \left( \ih S[x] - \ih S[y] \right) \ \rho (x_0, y_0)
\label{b3}
\end{equation}
The integral is over paths $x(t)$, $y(t)$ starting at $x_0$, $y_0$,
and both ending at the same final point $x_f$, where $x_f$, $x_0$
and $y_0$ are all integrated over, and weighted by the initial
state $\rho(x_0,y_0)$. The paths are also constrained to pass
through spatial gates at a sequence of times corresponding to the
projection operators.

However, the path integral representation of the decoherence
functional also points the way towards asking types of questions
that are not represented by homogeneous histories \cite{Har}.
In this article we are particularly interested in
the following question. Suppose a particle starts at $t=0$
in some quantum state. What is the
probability that the particle will either cross or never cross $x=0$
during the time interval $[0,\tau]$? In the path integral of the
form (\ref{b3}) it is clear
how to proceed. One sums over paths that, respectively,  either
always cross or never cross $x=0$ during the time interval.

How does this look in operator language?
The operator form of the decoherence functional is
\begin{equation}
D(\a, \a') = \Tr \left( C_{\a} \rho C_{\a'}^{\dag} \right)
\end{equation}
where
\begin{equation}
C_{\a} = P_{\a_n}(t_n) \cdots P_{\a_1} (t_1)
\end{equation}
The histories that never cross $x=0$
are represented by taking the projectors in $C_{\a}$
to be onto the positive $x$-axis, and then taking
the limit $ n \ria \infty$ and $t_k - t_{k-1} \ria 0 $.
The histories that always cross $x=0$ are then represented by
the object
\begin{equation}
{\bar C}_{\a} = 1 - C_{\a}
\label{b4}
\end{equation}
This is called an {\it inhomogenous} history, because it cannot be
represented as a single string of projectors.  It can however, be
represented as a {\it sum} of strings of projectors \cite{Har,Ish}.

The proper framework in which these operations, in particular (\ref{b4}),
are understood, is the so-called generalized quantum theory of
Hartle \cite{Har} and Isham {\it et al.} \cite{Ish}.
It is called ``generalized''
because it admits inhomogeneous histories as viable objects, whilst
standard quantum theory concerns itself entirely with homogeneous
histories. We will make essential use of inhomogeneous histories
in what follows.

In practice, for point particle systems,
decoherence is readily achieved by coupling to an environment.
Here, we will use the much studied case of the quantum Brownian
motion model, in which the particle is linearly coupled through
position to a bath of harmonic oscillators in a thermal state
at temperature $T$ and characterized by a dissipation coefficient
$\gamma$. The details of this model may be found elsewhere
\cite{CaL,FeV,Hal1,Hal2}.

We consider histories characterized only by the position of the
particle and the environmental coordinates are traced out.
The path integral representation of the decoherence functional
then has the form
\begin{equation}
D(\a, \a') = \int_{\a} {\cal D} x \ \int_{\a'} {\cal D} y
\ \exp \left( \ih S[x] - \ih S[y] + \ih W[x,y]
\right) \ \rho (x_0, y_0)
\end{equation}
where $W[x,y]$ is the Feynman--Vernon influence functional phase,
and is given by
\begin{equation}
W[x,y] = - m \gamma \int dt \ (x-y) ( \dot x + \dot y )
+ i { 2 m \gamma k T \over \hbar} \int dt \ (x-y)^2
\label{b5}
\end{equation}
The first term induces dissipation in the effective classical
equations of motion. The second term is responsible for thermal
fluctuations. It is also responsible for suppressing contributions
from paths $x(t)$ and $y(t)$ that differ widely, and produces
decoherence of configuration space histories.

The corresponding classical theory is no longer the mechanics of a
single point particle, but a point particle coupled to a heat bath.
The classical correspondence is now to a stochastic process which
may be described by either a Langevin equation, or by a
Fokker-Planck equation for a phase space probability distribution
$w(p,x,t)$:
\begin{equation}
{ \partial w \over \partial t}
= - { p \over m } { \partial w \over \partial x }
+ 2 \c { \partial ( p w ) \over \partial p }
+ D { \partial^2 w \over \partial p^2 }
\label{b7}
\end{equation}
where $w \ge 0 $ and
\begin{equation}
\int dp \ \int dx \ w(p,x,t) = 1
\end{equation}
When the mass is sufficiently large, this equation describes
near--deterministic evolution with small thermal fluctuations about it.

%\end{document}

%\documentstyle{article}
%\input defs.tex
%\begin{document}

\section{Spacetime Coarse Grainings}

We are generally interested in spacetime coarse grainings which
consist of asking for the probability that a particle does or does
not enter a certain region of space during a certain time interval.
However, the essentials of this question boil down to
the following simpler question: what is the probability
that the particle will either cross or not cross $x=0$ at any time
in the time interval $[0,\tau]$? We will concentrate on this question.

We briefly review the results of Yamada and Takagi \cite{YaT},
Hartle \cite{Har,Har2,Har3} and Micanek and Hartle \cite{MiH}. We
will compute the decoherence functional using the path integral
expression (\ref{b3}), which may be written
\begin{equation}
D(\a,\a') = \int dx_f \ \Psi^{\a}_\tau (x_f) \ \left( \Psi^{\a'}_\tau (x_f)
\right)^*
\end{equation}
where $\Psi^{\a}_\tau (x_f) $ denotes the amplitude obtained by summing
over paths ending at $x_f$ at time $\tau$,
consistent with the restriction $\a$ and consistent with
the given initial state, so we have
\begin{equation}
\Psi^{\a}_\tau (x_f) = \int_{\a}
{\cal D} x(t) \ \exp \left( \ih S[x] \right) \ \Psi_0 (x_0)
\end{equation}

Suppose the system starts out in the initial state $\Psi_0 (x) $
at $t=0$.
The amplitude for the particle to start in this initial state, and
end up at $x$ at time $\tau$, but without ever crossing
$ x = 0$, is
\begin{equation}
\Psi_\tau^r (x) = \int_{-\infty}^{\infty} dx_0 \ g_r (x,\tau | x_0, 0 ) \ \Psi_0 (x_0)
\label{c1}
\end{equation}
where $ g_r $ is the restricted Green function, {\it i.e.}, the sum
over paths that never cross $x=0$. For the free particle
considered here (and also for any system with a potential symmetric
about $x=0$), $g_r$ may be constructed by the method of images:
\begin{eqnarray}
g_r (x,\tau|x_0,0) &=& \left[ \theta (x) \ \theta (x_0 )
+ \theta (- x) \ \theta (- x_0) \right] \nonumber
\\
& \times & \left(g(x,\tau|x_0,0) - g(x,\tau|-x_0,0) \right)
\label{c2}
\end{eqnarray}
where $g(x,\tau |x_0,0)$ is the unrestricted propagator.

The amplitude to cross $x=0$ is
\begin{equation}
\Psi_\tau^c (x) = \int_{-\infty}^{\infty} dx_0 \ g_c (x,\tau | x_0, 0 ) \ \Psi_0 (x_0)
\end{equation}
where $g_c (x,\tau | x_0, 0 ) $ is the crossing propagator, {\it
i.e.}, the sum over paths which always cross $x=0$. This breaks up
into two parts. If $x$ and $x_0$ are on opposite sides of $x=0$,
it is clearly just the usual propagator $ g(x,\tau | x_0, 0 ) $. If
$x$ and $x_0$ are on the same side of $x=0$, it is given by $ g
(-x,\tau | x_0, 0) $. This may be seen by reflecting the segment of
the path after last crossing about $x=0$ \cite{HaO}.
(Alternatively, this is just the usual propagator minus the
restricted one). Hence,
\begin{eqnarray}
g_c (x,\tau | x_0, 0 ) &=& \left[ \theta
(x) \theta (-x_0 ) + \theta (-x) \theta ( x_0 ) \right] \ g (x,\tau |
x_0, 0 ) \nonumber \\
&+& \left[ \theta (x) \ \theta (x_0 ) + \theta (- x)
\theta (- x_0) \right] \ g (-x, \tau | x_0, 0 )
\end{eqnarray}
The crossing propagator may also be expressed in terms of the
so-called
path decomposition expansion, a form which is sometimes useful
\cite{AuK,vB,Hal3,HaO,ScZ}.

Inserting these expressions in the decoherence function, Yamada and
Takagi found that the consistency condition may be satisfied exactly
by states which are antisymmetric about $x=0$. The probability of
crossing $x=0$ is then $0$ and the probability of not crossing is
$1$. What is happening in this case is that the probability flux
across $ x=0 $, which clearly has non-zero components going both to
the left and the right, averages to zero.

Less trivial probabilities are obtained in the case where one asks
for the probability that the particle remains always in $x>0$ or
not, with an initial state with support along the entire $x$-axis
\cite{Har3}. The probabilities become trivial again, however, in
the interesting case of an initial state with support only in $x>0$.

Yamada and Takagi have also considered the case of the probability
of finding the particle in a spacetime region \cite{YaT}. That is, the
probability that the particle enters, or does not enter, the spatial
interval $\Delta$, at any time during the time interval $ [0,t]$.
Again the consistency condition is satisfied only for very special
initial states and the probabilities are then rather trivial.

In an attempt to assign probabilities for arbitrary initial states,
Micanek and Hartle considered the above
results in the limit that the time interval $[0,\tau]$ becomes very
small \cite{MiH}. Such an assignment must clearly be possible in
the limit $\tau \ria 0 $.
They found that both the off-diagonal terms of
the decoherence functional $D$ and the crossing probability $p$
are of order
$ \epsilon = \left( \hbar t / m \right)^{\half} $ for small $t$, and
the probability $\bar p$ for not crossing is of order $1$.
Hence $p + \bar p \approx 1 $.
They therefore argued that probabilities can be assigned if $t$
is sufficiently small.
On the other hand, we have the exact relation,
\begin{equation}
p + \bar p + 2 {\rm Re} D = 1
\end{equation}
$ {\rm Re} D $ represents the degree of fuzziness in the definition of
the probabilities. Since it is of the same order as $\bar p$, one may
wonder whether it is then valid to claim approximate consistency.
Another condition that may be relevant is the condition
\begin{equation}
|D|^2 < < p \bar p
\end{equation}
which was suggested in Ref. \cite{DoH} as a measure of
approximate decoherence, and is clearly satisfied in this case.

We conclude from these various studies that
for a system consisting of a single point particle,
crossing probabilities can be assigned to histories only in a
limited class of circumstances.

There is one particularly important case in which this lack of
probability assignment is perhaps unsettling. Consider a
wave packet that starts at  $x_0 > 0 $ moving towards the origin.
The amplitude for not crossing is given by the restricted
amplitude (\ref{c1}) and the restricted propagator (\ref{c2}).
However, in the case where the centre of the wave packet reaches
the origin during the time interval, it is easily seen from the
propagator (\ref{c2}) that after hitting the origin there is a
piece of the wave packet which is reflected back into $x>0$ (this
is the image wave packet that has come from $x<0$). This means that
we have the counterintuitive result that the probability for
remaining in $x>0$ is not in fact close to zero \cite{Har,Ya1}
as one would expect. It is unsettling because one sometimes thinks
of wave packets as being the closest thing quantum theory has to a
classical path, yet the behaviour of the wave packet in this case
is utterly different to the corresponding expected classical
behaviour.

Although counterintuitive, it is not that
disturbing, since with this initial state, the histories for
crossing and not crossing do not satisfy the consistency
condition, so we should not expect them to agree with our physical
intuition. Still, it would be reassuring to see that the formalism
set up so far yields the intuitively expected classical limit
under appropriate circumstances. To obtain that, we need a
decoherence mechanism, and this we now consider.

%\end{document}

%\documentstyle{article}
%\input defs.tex

%\begin{document}

\section{Decoherence of Spacetime Coarse-Grained Histories
in the Quantum Brownian Motion Model}

We have seen that crossing probabilities can only be assigned in
the decoherent histories approach for very special initial states,
and furthermore, we do not get an intuitively sensible classical
limit for wave packet initial states.
It is, however, well-known that most sets of histories of interest
do not in fact exhibit decoherence without the presence of some
physical mechanism to produce it. In this Section, we therefore
discuss a modified situation consisting of a point particle
coupled to a bath of harmonic oscillators in a thermal state. This
model, the quantum Brownian motion model \cite{QBM}, produces
decoherence of histories of positions in a variety of situations.

This explicit modification of the single particle system means
that the corresponding classical problem (to which the quantum
results should reduce under certain circumstances) is in fact a
stochastic process described by either a Langevin equation or by a
Fokker-Planck equation. It is therefore appropriate to first study
the crossing problem in the corresponding classical stochastic
process (see for example, Refs.\cite{Sie,BuT,BoD,MaW,Zaf}, and
references therein).

\subsection{The Crossing Time Problem in Classical Brownian Motion}

Classical Brownian motion may be described by the Fokker-Planck
equation (\ref{b7}) for the phase space probability distribution
$w(p,x,t)$. For simplicity we will work in the limit of negligible
dissipation, hence the equation is,
\begin{equation}
{ \partial w \over
\partial t} = - { p \over m } { \partial w \over \partial x} + D {
\partial^2 w \over \partial p^2 }
\label{d1}
\end{equation}
where $ D = 2 m \gamma k T $.
The Fokker-Planck equation is to be solved subject
to the initial condition
\begin{equation}
w(p,x,0) = w_0 (p,x)
\label{d2}
\end{equation}

Consider now the crossing time problem in classical Brownian
motion. The question is this. Suppose the initial state is localized
in the region $x>0$. What is the probability that, under evolution
according to the Fokker-Planck equation (\ref{d1}), the particle either
crosses or does not cross $x=0$ during the time interval $[0,\tau]$?

A useful way to formulate spacetime questions of this type
is in terms of the Fokker-Planck propagator,
$ K(p,x,\tau|p_0, x_0, 0 ) $.
The solution to (\ref{d1}) with the initial condition
(\ref{d2}) may be written in terms of $K$ as,
\begin{equation}
w (p,x,\tau) =
\int_{- \infty}^{\infty} dp \int_{- \infty}^{\infty} dx
\ K(p,x,\tau|p_0,x_0,0) \ w_0 (p,x)
\label{d3}
\end{equation}
The Fokker-Planck propagator satisfies the Fokker-Planck equation
(\ref{d1}) with respect to its final arguments, and satisfies
delta function initial conditions,
\begin{equation}
K(p,x,0|p_0,x_0,0) \ = \ \d (p-p_0) \ \d (x - x_0)
\label{d4}
\end{equation}
For the free particle without dissipation, it is given explicitly
by
\begin{eqnarray}
K(p,x,\tau|p_0,x_0,0) &=& N \ \exp  \left( - \a (p - p_0 )^2 -\b
(x- x_0 - {p_0 \tau \over  m} )^2 \right. \nonumber
\\
&+& \left. \epsilon (p-p_0 )(x-x_0 - {p_0 \tau \over  m}) \right)
\label{d5}
\end{eqnarray}
where $N$, $\a$, $\b$ and $\epsilon$ are given by
\begin{equation}
\a = { 1 \over D \tau}, \quad \b = { 3m^2 \over D\tau^3},
\quad \epsilon = { 3 m \over D \tau^2},
\quad N = \left( { 3 m^2 \over 4 \pi D^2 \tau^4} \right)^{\half}
\label{d6}
\end{equation}
(with $ D = 2 m \gamma k T $).
An important property it satisfies is the composition law
\begin{equation}
K (p,x,\tau |p_0,x_0,0) = \int_{-\infty}^{\infty} dp_1 \int_{-\infty}^{\infty}
dx_1 \ K(p,x,\tau|p_1,x_1,t_1) \ K(p_1,x_1,t_1 | p_0,x_0, 0 )
\label{d7}
\end{equation}
where $ \tau > t_1 > 0 $.

For our purposes, the utility of the Fokker-Planck propagator is
that it may be used to assign probabilities to individual paths in
phase space. Divide the time interval
$[0,\tau]$ into subintervals, $t_0 = 0,t_1,t_2, \cdots t_{n-1}, t_n = \tau $.
Then in the limit that the subintervals go to zero, and $n \ria \infty $
but with $\tau$ held constant, the quantity
\begin{equation}
\prod_{k=1}^n \ K(p_k, x_k, t_k | p_{k-1}, x_{k-1}, t_{k-1} )
\label{d8}
\end{equation}
is proportional to
the probability for a path in phase space. The probability for
various types of coarse grained paths (including spacetime coarse
grainings) can therefore be calculated by summing over this basic
object.

We are interested in the probability $w_r (p_n, x_n, \tau ) $
that the particle follows a path
which remains always in the region $ x> 0$ during the time interval
$ [0,\tau]$ and ends at the point $x_n > 0 $ with momentum $p_n$.
The desired total probabilities for crossing or not crossing can
then be constructed from this object. $w_r $
is clearly given by the continuum limit of the expression
\begin{eqnarray}
w_r (p_n, x_n,\tau) & = &
\int_0^{\infty} dx_{n-1} \cdots \int_0^{\infty} dx_1 \int_0^{\infty}
dx_0 \int_{-\infty}^{\infty} dp_{n-1} \cdots
\int_{-\infty}^{\infty} dp_1
\int_{-\infty}^{\infty} dp_0 \nonumber
\\
& \times & \ \prod_{k=1}^n \ K(p_k, x_k, t_k | p_{k-1}, x_{k-1}, t_{k-1} )
\ w_0 (p_0, x_0 )
\label{d9}
\end{eqnarray}

Now it is actually more useful to derive a differential equation and
boundary conditions for $w_r (p,x,\tau)$, rather than attempt to
evaluate the above multiple integral. First of all, it is clear
from the properties of the propagator that
$ w_r (p,x,\tau) $ satisfies the Fokker-Planck equation (\ref{d1}) and
the initial condition (\ref{d2}). However, we also expect some sort of
condition at $x=0$. From the explicit expression for the propagator
(\ref{d5}), (\ref{d6}), we see that in the continuum limit, the propagator between
$ p_{n-1}, x_{n-1} $ and the final point $p_n, x_n $ becomes proportional
to the delta function
\begin{equation}
\delta \left( x_n  -x_{n-1} - p_n \tau / m  \right)
\label{d10}
\end{equation}
Since $x_{n-1} \ge 0 $, when $x_n= 0$ this delta function will give
zero when $p_n > 0 $, but could be non-zero when $p_n < 0 $. Hence
we deduce that the boundary condition on $w_r (p,x,t) $ is
\begin{equation}
w_r (p,0,t) = 0, \quad {\rm if } \quad p> 0
\label{d11}
\end{equation}
This is the absorbing boundary condition usually given for the
crossing time problem \cite{MaW,WaU} (although this argument for
it does not seem to have appeared elsewhere).

It is now convenient to
introduce a restricted propagator $K_r (p,x,\tau|p_0,x_0,0)$, which
propagates $w_r (p,x,\tau) $. That is,
$K_r $ satisfies the delta function initial conditions (\ref{d11}) and the same
boundary conditions as $w_r$, Eq.(\ref{d11}). Since the original Fokker-Planck
equation is not invariant under $x \rightarrow - x$, we cannot
expect that a simple method of images (of the type used in Section
3), will readily yield the restricted propagator $K_r$.
$K_r$ has recently been found \cite{BoD},
using a modified
method of images technique due to Carslaw \cite{Car}, and we
briefly summarize those results.

Consider first the usual Fokker-Planck propagator (\ref{d5}).
Introducing the coordinates
\begin{eqnarray}
X & = &{p \over m } - { 3 x \over 2 \tau }, \quad Y = { \sqrt {3} x \over
2 \tau}
\\
X_0 & = & - { p_0 \over 2 m } - { 3 x_0 \over 2 \tau},
\quad Y_0 =  { \sqrt {3} \over 2 } \left( { p_0 \over m }
+ { x_0 \over \tau } \right)
\label{d12}
\end{eqnarray}
the propagator (\ref{d5}) becomes,
\begin{equation}
K = { \sqrt {3} \over 2 \pi \tilde t^2 }
\ \exp \left( - { (X- X_0 )^2 \over \tilde t }
- { (Y - Y_0 )^2 \over \tilde t } \right)
\label{d13}
\end{equation}
Here, $\tilde t = D \tau / m^2 $.
Now go to polar coordinates,
\begin{eqnarray}
X & = & r \cos \theta, \quad Y = r \sin \theta \\
X_0 & = & r' \cos \theta', \quad Y_0 = r' \sin \theta'
\label{d14}
\end{eqnarray}
Then from (\ref{d14}), it is possible to construct a so-called multiform
Green function \cite{Car},
\begin{equation}
g (r, \theta, r', \theta' ) =
{ \sqrt {3} \over 2 \pi^{3/2} \tilde t^2 }
\exp \left( - { r^2 + r'^2 - 2 r r' \cos (\theta - \theta')
\over \tilde t } \right) \ \int_{-\infty}^a d \lambda  \ e^{- \lambda^2}
\label{d15}
\end{equation}
where
\begin{equation}
a = 2 \left( { r r' \over \tilde t } \right)^{\half}
\ \cos \left( { \theta - \theta' \over 2} \right)
\label{d16}
\end{equation}
Like the original Fokker-Planck propagator, this object is a
solution to the Fokker-Planck equation with delta function initial
conditions, but differs in that it has the property that it is
defined on a two-sheeted Riemann surface and has period $ 4 \pi $.
The desired restricted propagator $K_r$ is then given by
\begin{equation}
K_r (p,x,\tau|p_0, x_0, 0 ) = g (r, \theta, r', \theta' )  -
g (r, \theta, r', - \theta' )
\label{d17}
\end{equation}
The point $x=0$ for $p>0$ is $\theta = 0 $ in the new coordinates,
and the above object indeed vanishes at $ \theta = 0$. Furthermore,
the second term in the above goes to zero at $\tau=0$, whilst the
first one goes to a delta function as required.

The probability of not crossing the surface during the time interval
$ [0,t]$ is then given by
\begin{equation}
p_r = \int_{-\infty}^{\infty} dp \int_0^{\infty} dx
\int_{-\infty}^{\infty} dp_0 \int_0^{\infty} dx_0
\ K_r (p,x,\tau | p_0, x_0, 0 ) \ w_0 ( p_0, x_0 )
\label{d18}
\end{equation}
The probability of crossing must then be $p_c = 1 - p_r $,
which can also be written,
\begin{equation}
p_c = \int_{-\infty}^0 dp
\int_{-\infty}^{\infty} dp_0 \int_0^{\infty} dx_0
\ { p \over m } \ K_r (p,x=0,\tau | p_0, x_0, 0 ) \ w_0 ( p_0, x_0 )
\label{d19}
\end{equation}
This completes the discussion of the classical stochastic problem.

\subsection{The Crossing Time Problem in Quantum Brownian Motion}

We now consider the analogous problem in the quantum case.
We therefore attempt to repeat the analysis of Section 3, but using
instead of (\ref{b7}), the decoherence functional appropriate to the
quantum Brownian motion model. It may be written,
\begin{equation}
D(\a, \a' ) = {\rm Tr} \left( \rho_{\a \a'} \right)
\label{d20}
\end{equation}
where
\begin{equation}
\rho_{\a \a'} (x_f, y_f) =
\int_{\a} {\cal D} x \int_{\a'} {\cal D} y \ \exp \left( \ih S[x]
- \ih S[y] + \ih W[x,y] \right) \ \rho_0 (x_0, y_0)
\label{d21}
\end{equation}
Here, $W[x,y]$ is the influence functional phase (\ref{a5}),
but with the
dissipation term neglected.
The sum is over all paths $x$, $y$ which are consistent with
the coarse graining $\a$, $\a'$, and end at the final points
$x_f, y_f$.

We will concentrate on the case in which the initial density operator has
support only on the positive axis, and we ask for the probability
that the particle either crosses or never crosses $x=0$ during the
time interval $[0,\tau]$. The history label $\a$ takes two values,
which we denote $\a=c$ and $\a=r$ for, respectively, crossing
and not crossing.

The objects $\rho_{\a \a'}$ defined in Eq.(\ref{d21})
actually obeys a master equation,
\begin{equation}
i \hbar { \partial  \rho \over \partial t} =
- { \hbar^2 \over 2m } \left( { \partial^2 \rho \over \partial x^2 }
- { \partial^2 \rho \over \partial y^2 } \right)
- \ih D (x-y)^2 \rho
\label{d21b}
\end{equation}
This is the usual master equation for the evolution of the density
operator of quantum Brownian motion \cite{CaL}.
The objects $\rho_{\a \a'}$ are then found by solving this equation
subject to matching the initial state $\rho_0$, and also to
the following boundary conditions (which follow from the path
integral representation):
\begin{eqnarray}
\rho_{rr} (x,y) &=& 0, \quad {\rm for} \quad x \le 0 \quad {\rm and}
\quad y \le 0
\\
\rho_{rc} (x,y) &=& 0, \quad {\rm for} \quad x \le 0
\\
\rho_{cr} (x,y) &=& 0, \quad {\rm for} \quad y \le 0
\end{eqnarray}
Given $\rho_{rr}, \rho_{rc}, \rho_{cr} $, the quantity $\rho_{cc}$
may be calculated from the relation,
\begin{equation}
\rho_{rr} + \rho_{rc} + \rho_{cr} + \rho_{cc} = \rho
\end{equation}

In the unitary case, this problem was solved very easily using the
method of images.
The problem in the non-unitary case treated here, however, is that
the master equation is {\it not} invariant under $x \ria -x $
(or under $ y \ria - y$), hence $\rho (-x,y)$ and $\rho(x,-y)$
are {\it not} solutions to the master equation. The method of images
is therefore not applicable in this case (contrary to the claim in
Ref.\cite{Har}). As far as an analytic
approach goes, this represent a very serious technical problem.
Restricted propagation problems are very hard to solve analytically
in the absence of the method of images. However, the presence
of the decohering environment allows for an approximate solution
of the problem. This is described in detail in Ref.\cite{HaZa}. The
results are intuitively clear and we summarize them here.

First of all, decoherence of position histories in this model
is extremely good, so $\rho_{rc} \approx  0 $, $\rho_{cr} \approx 0$.
We may therefore assign probabilities for not crossing and for
crossing, and these are equal respectively to
$ {\rm Tr} \rho_{rr} $ and $ {\rm Tr} \rho_{cc} $.
To see what these probabilities are, we make use of the Wigner
representation of the
density operator \cite{BaJ}:
\begin{equation}
W(p,x) = { 1 \over 2 \pi \hbar} \int_{-\infty}^{\infty}
d \xi \ e^{- \ih p \xi}
\ \rho( x + { \xi \over 2} , x - { \xi \over 2} )
\label{d22}
\end{equation}
The Wigner representation is very useful in studies of the
master equation, since it is similar to a classical phase
space distribution function. Indeed, for quantum Brownian
motion model with a free particle, the Wigner function obeys
the same Fokker-Planck equation (\ref{d1}) as the analogous
classical phase space distribution function. What makes
it fail to be a classical phase space distribution is that
it can take negative values. However, it can be shown that
the Wigner function becomes positive after a short time
(typically the decoherence time), and numerous authors
have discussed its use as an approximate classical phase
space distribution, under these conditions \cite{HaZo}.

Given approximate decoherence, it was shown at some length
in Ref.\cite{HaZa} using the path integral (\ref{d21})
that the Wigner transform of $\rho_{rr}$ is given by
\begin{equation}
W_{rr} ( m \dot X_f, X_f ) =
\int_r {\cal D} X \ \exp \left(
- { m \over 8 \gamma k T}
\int dt \ddot X^2 \right) \ W_0 (m \dot X_0, X_0 )
\label{d23}
\end{equation}
where the functional integral over $X(t)$ is over paths which lie in
$X>0$, and match $X_f$ and $\dot X_f $ at the final time.
If the paths $X(t)$ were not restricted, Eq.(\ref{d23}) would in
fact be a path integral representation of the Fokker-Planck
propagator (\ref{d5}) \cite{Kle}. With the restriction $X>0$,
it may be shown that it is a representation of the restricted
Fokker-Planck propagator (\ref{d17}) or (\ref{d9}).

It then follows that the probabilities for not crossing and for
crossing $ x = 0 $ are given, to a good approximation, by the
classical stochastic results (\ref{d18}), (\ref{d19}),
with the classical phase space distribution function $w_0$ replaced by
the initial Wigner function $W_0$ in the quantum case.
This result is the expected and intuitively obvious one, although
as outlined in Ref.\cite{HaZa}, it is a non-trivial matter to show
that the boundary conditions on $\rho_{\a \a'}$ in the quantum case
reduce to the boundary conditions on $W$ appropriate to the classical
stochastic problem.

\subsection{Properties of the Solution}

Some simple
properties of our results may be seen by examining the path integral
form of the solution (\ref{d23}). The important case
to consider is the motion of a wavepacket, since this is the
situation that gave problematic results in Section 3.
We take an initial state consisting of a wavepacket concentrated at
some $ x> 0 $, and moving towards the origin. We are interested in
the probability of whether it will cross $x=0$ or not during some
time interval, under the evolution by the path integral
(\ref{d23}).

The integrand in (\ref{d23}) is peaked about the unique path for which
$\ddot X = 0$ with the prescribed values of $X_0$ and $ \dot X_0 $.
This is of course the classical path with the prescribed initial
data. From (\ref{d23}), the spatial width
$( \Delta X )^2$ of the peak is of order $ \gamma k T
/ ( m \tau^3) $.
If the classical path does not cross $x=0$ and approaches $x=0$ no
closer than a distance $\Delta X $ during the time interval, then it will
lie well within the integration range $X>0$, and the propagation is
essentially the same as unrestricted propagation, since the dominant
contribution to the integral comes from the region $ X>0 $. It is
then easy to see, from the normalization of the Wigner function,
that the probability of not crossing is approximately $1$, the
intuitively expected result.

If the classical path crosses $x=0$ during the time interval, it
will lie outside the integration range of $X$ for time slices
after the time at which it crossed. If it crosses sufficiently early
that an entire wave packet packet of width $\Delta X$ may enter
$x<0$ before time $\tau $, then the functional integration will
sample only the exponentially small tail of the integrand, so
$W_{rr}$ will be very small. The probability of not crossing will
therefore be close to zero, again the intuitively expected result.

The inclusion of the environment therefore restores the intuitively
sensible classical limit to the quantum case of Section 3.

In the above simple examples, the crossing probabilities are
independent of the details of the environment, to a leading
order approximation.
It is clear that in a more precise expression, the crossing
probabilities will in fact depend on the features of the environment
({\it e.g.}, its temperature). One might find this slightly
unsettling, at least in comparison to quantum-mechanical
probabilities at a fixed moment of time, which depend only on the
state at that time and not on the details of how the property in
question might be measured. This is in keeping with an opinion
sometimes expressed on questions of time
in quantum mechanics -- that to specify times one has to specify
the physical mechanism by which it is measured \cite{Lan}.

%\end{document}

%\documentstyle{article}

%\input defs.tex

%\begin{document}

\section{A Detector Model}

Although the results of the previous sections produced
mathematically viable candidates for the probabilities of crossing
and not crossing $x=0$, it is by no means clear how they
correspond to a particular type of measurement. As noted in Section 2,
general theorems
exist showing that decoherence of histories implies the existence
at the final end of the histories of a record storing the
information about the decohered histories \cite{GeH2,GeH4}. This
means that there is {\it some} quantity at a fixed moment of time
which is correlated with the property of crossing or not crossing
$x=0$ during the time interval $[0,\tau]$, and which could in
principle be measured. Records associated with decoherence have,
however, been explicitly found only in a few simple cases (see
Ref.\cite{Hal5}, for example). For these reasons, it is of
interest to compare the approaches involving the decoherent
histories aproach with a completely different approach involving a
specific model of a detector.

We therefore introduce, following Ref.\cite{Hal6},
a model detector which is coupled to the
particle in the region $ x<0$, and such that it undergoes a
transition when the coupling is switched on.  Such detectors have
certainly been considered before (see, {\it e.g.},
Ref.\cite{AOP}). The particle could, for example, be coupled to a
simple two-level system that flips from one level to the other when
the particle is detected. One of the difficulties of many detector
models, however, is that if they are modeled by unitary quantum
mechanics, the possibility of the reverse transition exists. Because
quantum mechanics is fundamentally reversible, the detector could
return to the undetected state under its self-dynamics,  even when
the particle has interacted with it.

To get around this difficulty, we appeal to the fact that realistic
detectors have a very large number of degrees of freedom, and are
therefore effectively {\it irreversible}. They are designed so that
there is an overwhelming large probability for them to make a
transition in one direction rather than its reverse.   We consider a
simple model detector that has this property.  This is achieved by
coupling a two-level system detector to a large environment, which
makes its evolution effectively irreversible.

The detector is a two-level system, with levels $ | 1 \ra
$ and $ | 0 \ra $, representing the states of no detection and
detection, respectively. Introduce the raising and lowering
operators
\begin{equation}
\sp = | 1 \ra \la 0 |, \quad \sm = | 0 \ra \la 1 |
\end{equation}
and let the Hamiltonian of the detector be $H_d = \half \hbar
\omega \s_z $, where
\begin{equation}
\sigma_z = | 1 \ra \la 1 | - | 0 \ra \la 0 |
\end{equation}
so $ | 0 \ra $ and $ | 1 \ra $ are eigenstates of $H_d$ with
eigenvalues $ - \half \hbar \omega $ and $ \half \hbar \omega $
resepectively. We would like to couple the detector to a free
particle in such a way that the detector makes an essentially
irreversible transition from $ | 1 \ra $ to $ | 0 \ra $ if the
particle enters $x < 0 $, and remains in $ | 1 \ra $ otherwise. This
can be arranged by coupling the detector to a large environment of
oscillators in their ground state, with a coupling proportional to $
\theta (-x)$. This means that if the particle enters the region
$x<0$,  the detector becomes coupled to the large environment
causing it to undergo a transition. Since the environment is in its
ground state, if the detector initial state is the higher energy
state $ | 1 \ra $ it will, with overwhelming probability, make a
transition from $ | 1\ra $ to the lower energy state $ | 0 \ra $.
A possible Hamiltonian describing this
process for the three-component system is
\begin{equation}
H = H_s + H_d + H_{\E} + V(x)  H_{d \E }
\end{equation}
where the first three terms are the  Hamiltonians of the particle,
detector and environment respectively, and $H_{d \E}$ is the
interaction Hamiltonian of the detector and its environment.
The simplest choice
of environment is a collection of harmonic oscillators,
\begin{equation}
H_{\E} = \sum_n \hbar \omega_n a_n^{\dag} a_n
\end{equation}
and we take the coupling to the detector to be via the interaction
\begin{equation}
H_{d \E} = \sum_n \hbar \left( \kappa^*_n \sm a_n^{\dag} +
\kappa_n \sp a_n \right)
\end{equation}
An environment consisting of an electromagnetic field, for example,
would give terms of this general form.
$V(x)$
is a potential concentrated in $x<0$ (and we will eventually make
the simplest choice, $ V(x) = \theta (-x) $, but for the moment we
keep it more general). The important feature is that the interaction
between the detector and its environment, causing the detector to
undergo a transition, is switched on only when the particle is in
$x<0$.

A similar although more
elaborate model particle detector has been previously studied by
Schulman \cite{Sch1} (see also Refs.\cite{Sch2}). The advantage
of the present model is that it is easier to solve explicitly.

We are interested in the reduced dynamics of the particle and
detector with the environment traced out. Hence we seek a master
equation for the reduced density operator $\rho$ of the particle and
detector. With the above choices for $H_{\E}$ and $H_{d \E} $, the
derivation of the master equation is standard \cite{Carm,Gar} and
will not be repeated here. There is the small complication of the
factor of $V(x)$ in the interaction term, but this is readily
accommodated. We assume a factored initial state, and we assume that
the environment starts out in the ground state. In a Markovian
approximation (essentially the assumption that the environment
dynamics is much faster than detector or particle dynamics), and in
the approximation of weak detector-environment coupling, the master
equation is
\begin{equation}
\dot \rho = - \ih [ H_s + H_d, \rho]
- { \gamma \over 2} \left( V^2 (x ) \sp \sm \rho \ +  \rho
\sp \sm  V^2 (x)  \ -  \ 2 V (x) \sm \rho \sp V (x )
\right)
\label{e1}
\end{equation}
Here, $\gamma$ is a phenemonological constant determined  by the
distribution of oscillators in the environment and underlying
coupling constants. The frequency $\omega $ in $H_d$ is also
renormalized to a new value $\omega'$.

Eq.(\ref{e1}) is the sought-after description of a particle coupled to an
effectively irreversible detector in the region $ x< 0 $. In the
dynamics of the detector plus environment only ({\it i.e.}, with
$V=1$ and $H_s=0$), it is readily shown that every initial state
tends to the state $ | 0 \ra \la 0 | $ on a timescale $\gamma^{-1}$.
With the particle coupled in, if the initial state of
the detector is chosen to be $ | 1 \ra \la 1 | $, it undergoes an
irreversible transition to the state $ | 0 \ra \la 0 | $ if the
particle enters $ x < 0 $, and remains in its initial state
otherwise.

Eq.(\ref{e1}) is in fact of the Lindblad form (the most general
Markovian master equation preserving density operator properties
\cite{Lin}). A similar detection scheme based on a postulated master
equation similar to (\ref{e1})) was previously considered in
Ref.\cite{Jad}.

The master equation (\ref{e1}) is easily solved by writing
\begin{equation}
\rho  =
\rho_{11} \otimes | 1 \ra \la 1 |
+ \rho_{01} \otimes | 0 \ra \la 1 |  +
\rho_{10} \otimes | 1 \ra \la 0 |
+ \rho_{00} \otimes | 0 \ra \la 0 |
\end{equation}
We suppose that the particle starts out in an initial state $ |
\Psi_0 \ra $, hence the master equation is to be solved subject to
the initial condition,
\begin{equation}
\rho ( 0 ) = | \Psi_0 \ra \la \Psi_0 | \otimes | 1 \ra \la 1 |
\end{equation}
The probability that the detector does not register during
$[0,\tau]$ is
\begin{equation}
p_{nd} = {\rm Tr} \rho_{11} = \int_{- \infty}^{\infty} dx
\ \rho_{11}(x,x, \tau )
\end{equation}
and the probability that it registers is
\begin{equation}
p_d = {\rm Tr} \rho_{00} = \int_{- \infty}^{\infty} dx
\ \rho_{00}(x,x, \tau )
\end{equation}
(where the trace is over the particle Hilbert space).
Clearly $p_{nd} + p_d = 1 $, since $ {\rm Tr} \rho = 1 $.

The explicit solution to the master equation is straightforward
and was carried out
in Ref.\cite{Hal6}. There, it was shown that, when $V(x) = \theta (-x)$,
the solution for $\rho_{11}$ may be written
\begin{equation}
\rho_{11} (t) = \exp \left( - \ih H_s t - { \gamma \over 2} V t \right)
\ \rho_{11} (0)
\ \exp \left( \ih H_s t - { \gamma \over 2} V  t \right)
\label{e2}
\end{equation}
What is particularly interesting about this expression is that it
can be factored into a pure state. Let
$ \rho_{11} = | \Psi \ra \la \Psi | $. Then, noting that
$ \rho_{11} (0) = | \Psi_0 \ra \la \Psi_0 | $,
Eq.(\ref{e2}) is equivalent to
\begin{equation}
   | \Psi (t) \ra = \exp \left( - \ih H_s t - { \gamma \over 2} V t \right)
   | \Psi_0 \ra
\label{e3}
\end{equation}
The probability for no detection is then
\begin{equation}
    p_{nd}
    = \int_{-\infty}^{\infty} dx \ | \Psi (x, \tau ) |^2
\label{e4}
\end{equation}
The pure state (\ref{e3}) evolves according to a Schr\"odinger
equation with an imaginary contribution to the potential, $ -
\half i \hbar \gamma V $. Complex potentials of precisely this
type have been used previously in studies of arrival times, as
phenomenological devices, to imitate absorbing boundary conditions
(see, for example Refs.\cite{All,MBM,PMB}). Here, the appearance
of a complex potential is {\it derived} from the master equation
of a particle coupled to an irreversible detector, which in turn
may be derived from the unitary dynamics of the combined
particle--detector--environment system.

In summary, this detector model nicely reproduces earlier
phenomenological results on arrival times. In Ref.\cite{MuL}
it is also shown that the expression (\ref{e3}), (\ref{e4}),
is very closely related to the ``ideal'' arrival time
distribution of Kijowski \cite{Kij}. An improved more physically
realistic irreversible detector model (although more difficult to solve
analytically) was recently put forward by Muga et al. \cite{Muga}.

%\end{document}

%\documentstyle{article}

%\input defs.tex

%\begin{document}

\section{A Comparison of the Decoherent Histories Result
with the Detector Result}

We may now compare the two candidate expressions for the crossing
time probabilities, one from decoherent histories with an
environment, the other from an irreversible detector model.
We will quickly see that the two results are not in fact
very close, but it is perhaps of interest to see exactly
why, and how they may be improved.

We first massage the decoherent histories result into a more
suitable form. Consider the probability for remaining in $x>0$.
From (\ref{d21}) it is given by
\begin{eqnarray}
p_r &=& \int_r \D x (t) \int_r \D y (t)  \exp \left(
\ih S [ x(t) ] - \ih S [ y (t) ] \right)
\nonumber \\
&\times & \  \exp \left( - a \int dt \ (x-y)^2 \right) \ \rho_0
(x_0, y_0 ) \label{f1}
\end{eqnarray}
where $a = D/\hbar^2$. Following Ref.\cite{HaZa}, we make the
observation that the last exponential may be deconvolved:
\begin{equation}
\exp \left( - a \int dt \ (x-y)^2 \right) = \int {\cal D} \bx \
\exp \left( - 2 a \int dt \ ( x - \bx )^2 - 2 a \int dt \ (y - \bx
)^2 \right)
\end{equation}
Hence, assuming a pure initial state,
the probability (\ref{f1}) may be written,
\begin{eqnarray}
p_r &=& \int \D \bx (t) \int_r \D x(t) \ \exp \left( \ih S [x(t)]
- 2 a \int dt \ (x - \bx)^2 \right) \Psi_0 (x_0 )
\nonumber \\
&\times & \int_r \D y(t) \ \exp \left(- \ih S [y(t)] - 2 a \int dt
\ (y - \bx)^2 \right) \Psi_0^* (y_0 )
\end{eqnarray}
In these integrals, $\bx (t) $ is integrated over an infinite range,
but $x (t) $ and $y(t)$ are integrated only over the positive  real
line. This restriction is quite difficult to implement in practice
\cite{HaZa}.
However, because of the exponential factors, negative values of
$ \bx (t) $ are strongly suppressed, so we may take its range to be
over positive values only, with exponentially small error.
Furthermore, having done this we may then (for technical simplicity)
allow the range of $x(t)$ and $y(t)$ to be over the entire real
line, again with exponentially small error. Therefore, we
have that
\begin{eqnarray}
p_r & \approx & \int_r \D \bx (t)
 \int \D x(t) \ \exp \left( \ih S [x(t)] - 2 a \int dt \ (x - \bx)^2
\right) \Psi_0 (x_0 )
\nonumber \\
& \times & \int \D y(t) \ \exp \left(- \ih S [y(t)] - 2 a \int dt \
(y - \bx)^2 \right) \Psi_0^* (y_0 )
\end{eqnarray}
This may finally be written,
\begin{equation}
p_r \approx \int_r \D \bx (t) \ \la \Psi_{\bx} | \Psi_{\bx} \ra
\label{f2}
\end{equation}
where
\begin{equation}
\Psi_{\bx} (x_f, \tau ) = \int \D x(t) \ \exp \left( \ih S [x(t)]
- 2 a \int dt \ (x - \bx)^2 \right) \Psi_0 (x_0 ) \label{f3}
\end{equation}

Written in this way the probability has a natural interpretation in
terms of continuous quantum measurement. Eq.(\ref{f3}) is the wave function for
a system undergoing continuous measurement of its position along a
trajectory $\bx (t)$ to within a precision proportional to $ a^{-
\half}$. The probability for any such trajectory is $ \la \Psi_{\bx}
| \Psi_{\bx} \ra $, hence the probability to remain in the region
$x>0$ is obtained by integrating over $ \bx (t) > 0 $. The
probability (\ref{f1}), derived from the decoherent histories
approach, is therefore, to an excellent approximation, the same as the
result naturally obtained from continuous quantum measurement theory

Now we compare with the detector model.
The probability for no detection is computed from the
wave function (\ref{e3}). In a path integral representation,
this may be written,
\begin{equation}
\Psi_{nd} (x_, \tau )
= \int {\cal D} x (t)
\exp \left( \ih S [ x (t) ] -
{\gamma \over 2} \int_0^{\tau} dt \ V ( x (t) ) \right)
\Psi_0 (x_0)
\label{f4}
\end{equation}
The sum is over all paths $x(t)$ connecting $x_0$ at $t=0$ to
$x_f$ at $t=\tau$. The probability for no detection is then
quite simply
\begin{equation}
p_{nd} = \la \Psi_{nd} | \Psi_{nd} \ra
\label{f5}
\end{equation}
Whilst the two different expressions, (\ref{f2}), (\ref{f3}), versus
(\ref{f4}), (\ref{f5}) are similar in some ways, they are not
obviously close and suffer from a rather key difference. Eq.(\ref{f2})
is obtained by summing the probability for any path $\bx (t)$
over positive values of $\bx$. In Eqs.(\ref{f4}), (\ref{f5}),
by contrast, the restriction to paths in $x>0$
is already imposed in the amplitude. The difference between the
probabilities provided by the detector and those provided by the
decoherent histories approach is,
therefore, the difference between summing amplitudes and squaring, versus
squaring and then summing.

In the decoherent histories approach, the coupling to the
environment produces an effective measurement of the system that
is much finer than is required for the crossing time problem. It
effectively measures the entire trajectory, which is clearly much
more information than is required to determine whether or not the
particle enters $x<0$. In this sense this particular decoherent
histories model is much cruder than the detector model, since it
destroys far more interference than it really needs to in order
to define the crossing time. This is due to the form of the
particle-environment coupling which is linear in the particle's
position. It would be of interest to explore a decoherent
histories model with a more refined type of coupling which is more
specifically geared to the crossing time problem.

It is of interest to note that continuous quantum measurement
theory in fact suggests another candidate expression for the
probability of not crossing which is closer to the detector model.
Suppose that {\it before} squaring, we sum the amplitude
(\ref{f3}) over positive $\bx (t)$:
\begin{eqnarray}
\Psi_{+} (x_f, \tau ) & =
& \int \D x(t) \ \exp \left( \ih S [x(t)] \right)
\nonumber \\
& \times & \int_r \D \bx (t) \left( - 2 a \int dt \ (x - \bx)^2
\right) \Psi_0 (x_0 ) \label{f6}
\end{eqnarray}
The probability is then
\begin{equation}
p_+ = \la \Psi_+ | \Psi_+ \ra
\end{equation}
This expression for the probability of not entering $x<0$ is
completely natural from the point of view of continuous quantum
measurement theory. It does not follow from either the detector
model or from the decoherent histories approach presented here,
but one can regard it as yet another {\it proposal} with which to
define the arrival time probability. The amplitude (\ref{f6}) is
now more closely analogous to the detector result (\ref{f4}). To
see this, introduce the effective potential $V_{eff} (x) $ defined
by
\begin{equation}
\exp \left( - \int dt \ V_{eff} (x(t) ) \right) = \int_r \D \bx
(t) \left( - 2 a \int dt \ (x - \bx)^2 \right)
\end{equation}
The integral can be evaluated exactly, but it is clear that $
V_{eff} (x) \sim 0 $ for $x>>0$, and  $ V_{eff} (x) \sim 2 a x^2 $
for $x<< 0 $. Eq.(\ref{f6}) therefore has the same general form as
(\ref{f4}). The potential is not exactly the same, but has the
same physical effect, which is to suppress paths in $x<0$.

%\end{document}

%\documentstyle{article}

%\input defs.tex

%\begin{document}

\section{Timeless Questions in Quantum Theory}

We now briefly consider a related question in quantum theory that
involves time in a non-trivial way, which is in fact more closely
related to the Wheeler-DeWitt equation of quantum cosmology,
(\ref{a1}). This equation may be thought of as the statement that
the wave function of the system is in an energy eigenstate. As
stated in the Introduction, the equation contains no notion of
time, and indeed ``time'' and the notion of trajectories are
thought to somehow emerge from the wave function. To test this
idea, and hence to provide some sort of interpretation for the
wave function, we need to find an answer to the question, ``What
is the probability associated with a given region $\Delta$ of
configuration space when the system is in an energy eigenstate,
without any reference to time?".

Classically, the question is well-defined. A system with fixed
energy consists of a set of classical trajectories, perhaps with
some probability distribution on them. The classical trajectories
are just curves in configuration space, and the question is then
quite simply one of determining whether or not these curves
intersect the given region $\Delta $. But, like the arrival time
problem in non-relativistic quantum mechanics, the problem is
considerably harder to phrase in quantum theory.

To see the beginnings of the difficulties, we briefly consider
the following simple question for a two-dimensional system with
coordinates $x_1, x_2$: given that the system is in an energy
eigenstate, what is the value of $x_1$ given the value of $x_2$?
Slightly rephrased, what is the probability that the system
intersects the surface $x_2 = constant$ between $x_1$ and $x_1 + d
x_1$, at any time? An operator approach to the problem, for
example, takes the following form. For a free particle, the
classical trajectories are
\begin{equation}
x_1 (t)  = x_1 + { p_1 t \over m}, \quad x_2 (t)  = x_2 + { p_2 t
\over m}
\end{equation}
and we may eliminate $t$ between them to write,
\begin{equation}
x_1(t) = x_1 + { p_1 \over p_2} ( x_2 (t) - x_2 )
\label{g1}
\end{equation}
This is the classical answer to the question, what is the value of
$ x_1$ at a given value of $x_2$? One may attempt to raise this to
the status of an operator in the quantum theory. It commutes with
the free particle Hamiltonian,
\begin{equation}
H = \half (p_1^2 + p_2^2 )
\end{equation}
so is in this sense an observable of the theory -- measuring it
will not displace the system from an energy eigenstate of $H$.
This approach encounters problems, however, in defining (\ref{g1}).
It cannot be
made into a self-adjoint operator, due to the presence of the
$1/p_2$ factor. In this way it is very similar to the problem
of defining a time operator.

We will not pursue this approach any further here. Instead
we briefly report on two other approaches, which, exactly like
the approaches described in this article, use decoherent
histories, or a detector model.

The decoherent histories approach to the question involves summing
over paths in configuration space which either enter or do not enter
a given region $\Delta $ at {\it any} moment of time. In practice
this is achieved by summing over paths which either enter or do not
enter during a fixed time interval $[0,\tau]$, and then summing $\tau$
over an infinite range. The detailed construction of this is described
in Ref.\cite{HaTh}. As in the crossing time problem
described in Section 4, a decohering
environment is required to make the probabilities well-defined, and
we then expect the final result to be a reasonably simple formula
involving the Wigner function, closely analagous to the classical case.
The full details of this have yet to be worked out, but is is perhaps
useful to give here the classical result (which, although well-defined,
is not totally trivial).

We consider a $2n$-dimensional phase space with coordinates $\p, \x$.
Denote the classical trajectories by $\x^{cl}( t) $, and suppose
that they match the initial data $\p_0, \x_0$ at some fiducial
initial point $t=t_0$ (which is arbitrary). For a free particle,
\begin{equation}
\x^{cl} (t) = \x_0 + { \p_0  \over m} (t-t_0)
\end{equation}
Let $f_{\Delta} (\x ) $ be a characteristic function for the
region $\Delta $ so is $1$ inside $\Delta $ and zero outside.
We suppose that the classical system is described by a
phase space distribution function $w (\p, \x)$. To be
a true analogue of an energy eigenstate in the quantum case, $w$
has to be stationary, so
\begin{equation}
w (\p(t), \x(t) ) = w (\p (t+t_1), \x (t+t_1))
\label{g2}
\end{equation}
for any $t_1$.

We may now write down the probability for a classical trajectory
entering the region $\Delta $. It is,
\begin{equation}
p_{\Delta} = \int d^n \p_0 d^n \x_0 \ w (\p, \x ) \ \theta \left(
\int_{-\infty}^{\infty} dt \ f_{\Delta} (\x^{cl} (t) ) - \e \right)
\label{g3}
\end{equation}
Here, $\e$ is a small parameter which is taken to zero through
positive values, and is present to avoid ambiguities in the $\theta$
function at zero argument. The integral inside the $\theta$ function
is the total time spent by the trajectory $\x^{cl} (t)$ inside the
region $\Delta$, but we are only interested in whether this
time is positive or zero. The initial data $\p_0, \x_0$ are therefore
effectively integrated only over values for which the trajectory spends
a time in excess of $\e$ in the region $\Delta$. It is easy to see
that the whole construction is invariant under shifting the fiducial
point $t_0$. This is the analogue of reparametrization invariance
(or more generally, diffeomorphism invariance) in the Wheeler-DeWitt equation
Eq.(\ref{a1}). It is expected that a decoherent histories analysis
will yield a result of the approximate form (\ref{g3}) (with $w$ replaced
by the Wigner function).

The other approach to the question posed at the beginning of this
section is to use a detector model
(this is described in detail in Ref.\cite{Hal7}).
The detector model arises from Barbour's observation \cite{Bar}
that a
substantial insight into the Wheeler-DeWitt equation may be found
in Mott's 1929 analysis of alpha-particle tracks in a Wilson
cloud chamber \cite{Mot}. Mott's paper concerned the question of how the
alpha-particle's outgoing spherical wave state, $ e^{ikR}/R $,
could lead to straight line tracks in a cloud chamber. His
explanation was to model the cloud chamber as a collection of
atoms that may be ionized by the passage of the alpha-particle.
They therefore act as detectors that measure the
alpha-particle's trajectory. The probability that certain atoms
are ionized is indeed found to be strongly peaked when the atoms
lie along a straight line through the point of origin of the
alpha-particle.

Mott had in mind a time-evolving process, but
he actually solved the time-independent equation
\begin{equation}
\left( H_0 + H_d + \lambda H_{int} \right) | \Psi \ra = E | \Psi \ra
\end{equation}
Here $H_0$ is the alpha-particle Hamiltonian, $H_d$ is the
Hamiltonian for the ionizing atoms, and $H_{int}$ describes the
Coulomb interaction between the alpha-particle and the ionizing
atoms (where $\lambda$ is a small coupling constant). Now the
interesting point, as Barbour notes, is that Mott derived all the
physics from this equation with little reference to time. Mott's
calculation is therefore an excellent model for many aspects of
the Wheeler-DeWitt equation. In Ref.\cite{Hal7} a model of this
type is considered with a series of detectors, and it is shown
how to produce a plausible formula for the probability that
the system enters a series of regions in configuration space
without reference to time. A comparison of this approach with
the anticipated decoherent histories result (\ref{g3}) is
yet to be carried out.

%\end{document}

%\documentstyle{article}

%\input defs.tex

%\begin{document}

\section{Discussion}

We have reviewed a number of approaches to the crossing time
problem in non-relativistic quantum theory, primarily using the
decoherent histories approach. We have also briefly reviewed some
attempts to extend these ideas to models more closely related to
the Wheeler-DeWitt equation. On the face of it, the decoherent
histories approach appears to be particularly well adapted to this
problem, since it naturally incorporates the notion of trajectory,
and hence readily accommodates questions of a non-trivial temporal
nature. Having said that, however, good expressions for the
crossing time probability are not acquired very easily.

As described in Section 3, the decoherence or consistency
conditions are satisfied only for very special classes of initial
states. For a system consisting of a single point particle,
therefore, the decoherent histories approach does not supply an
answer to the crossing time problem for {\it arbitrary} initial
states. This is rectified by the inclusion of a thermal
environment, as described in Section 4, and probabilities for the
crossing time can then be obtained for arbitrary initial system
states. They do, however, depend to some extent on the environment
producing the decoherence, and moreover, they are essentially the
same as the classical stochastic results. One might therefore
criticize this result on the grounds that it is ``not very
quantum''. This is largely true, but the essential achievement of
Section 4 is to show that the decoherent histories approach can be
made to give the anticipated classical result. This was not true
of the earlier approaches reviewed in Section 3.

In Section 5, a detector model was introduced to give an
alternative expression for the crossing time probability, for the
purposes of comparison with the decoherent histories result. The
detector model gave a better result, in that it agreed and
substantiated an earlier result of Allcock \cite{All}, which in
turn is closely related to the ideal distribution of Kijowski
\cite{Kij}.

On comparison with the decoherent histories result, in Section 6,
it was easy to see that the environment in Section 4 produced far
more decoherence than is necessary to define the arrival time, and
in that sense, that particular environment is a very crude model
for the measurement of time. The comparison did, however, inspire
the proposal of a third candidate expression from which the
arrival time probability could be calculated, namely
Eq.(\ref{f6}), which is based on continuous quantum measurement
theory. This expression does not seem to have been considered
previously and will be explored in more detail elsewhere.

One might be led from these results to a somewhat negative
assessment of the decoherent histories approach's ability to
provide the crossing time probability. The somewhat crude nature
of the results of Section 4, are however, due to the choice of a
rather indiscriminate system-environment coupling, which
effectively measures the entire trajectory. It seems likely that a
much-improved result could be obtained through choice of a more
refined coupling better suited to this particular problem.

Furthermore, there is another aspect to the decoherent histories
approach in this context which has not yet been explored. Many
approaches to the arrival time problem are based on model
measuring devices, that is, physical systems in which one of the
dynamical variables is correlated with time in some way. The
detector model of Section 5 was of this type: one could think of
the two-state system as being some kind of clock or detector
attached to the particle, which switches on when the particle
enters the region $x<0$. By physically measuring the two-state
system at the end of the time interval $[0, \tau]$ of interest,
one expects to be able to deduce that the particle was in $x<0$,
or not, during the time interval. The outstanding question,
however, is this: how do we really know that the detector state is
correlated with whether or not the particle entered $x<0$?

This is where the decoherent histories approach comes in. We
consider a system consisting of the particle and a detector (and
possibly also an environment, if necessary). We then look at
histories in which both the final state of the detector and the
particle alternatives (whether or not it entered $x<0$ during
$[0,\tau]$) are specified. If these histories are decoherent, we
then obtain a joint probability distribution for the histories of
the particle and the final state of the detector, and we can ask
to what degree these two things are correlated. If they are
perfectly correlated, then the detector probability is exactly the
same as the probability of the detector and the particle
alternatives.

In brief, therefore, the decoherent histories approach will be a
useful tool in assessing the extent to which a proposed detector
really does its job \cite{Har1}. Many model detectors are proposed
essentially on the basis of classical arguments, but the
decoherent histories approach allows their effectiveness to be
checked in a genuinely quantum way. This possibility does not
appear to have been explored in the context of arrival times, but
will be considered elsewhere.

\subsection*{Acknowledgements}

I am extremely grateful to Gonzalo Muga for inviting me to
contribute to this volume. I would also like to thank Gonzalo for
inviting me to take part in the meeting {\it Time in Quantum
Mechanics} at La Laguna, Tenerife, in May 1998, which stimulated
a substantial proportion of the work described here.

\end{document}